\documentclass[aps,preprint]{revtex4-1}
\usepackage{amsmath}
\usepackage{graphicx}
\usepackage{epstopdf}

\begin{document}

\title{Rare Radiative Decays of Vector and Axial-Vector $B_{c}$ Mesons to $ (D_{s}, D^{*}_{s}, D_{s1}) $ Final States}

\author{ A. R. Olamaei\footnote{e-mail: olamaei@jahromu.ac.ir} }

\affiliation{ Physics Department, Jahrom University, Jahrom, P. O. Box 74137-66171, Iran }

\begin{abstract}
In this work, we use the QCD sum rule method to study the radiative decays of the vector and axial-vector $B_c$ mesons to each of three charmed strange mesons, $D_s$, $D^*_s$ and $D_{s1}$, through their dominant weak annihilation channels. We calculate all relevant  transition form factors, which are used to estimate the branching fractions at different channels. The order of branching ratios are obtained to be in the order of $ 10^{-6} - 10^{-5} $, which may be checked via differnet experiments.
~~~PACS numbers: 11.55.Hx, 13.20.-v, 13.20.He
\end{abstract}

\maketitle

\section{Introduction}

The mesons containing heavy quarks are rich factories to investigate new features of the Standard Model and beyond. Among them, $B_c$ meson is one of the most favorite candidates, because it is the only heavy meson consisting of two heavy quarks with different flavores.The pseudoscalar $B_c$ meson has been discovered in 1988 \cite{Abe:1998fb} and studied widely via different methods \cite{GI,EFG,GJ,ZVR,Fulc,GKLT,EQ,Kiselev04,IS2004,BV2000,Penin2004,Latt,Latt0909}. Indeed different decay channels of the pseudo-scalar $B_c$ meson have been studied widely via QCD sum rule method \cite{Aliev1,Aliev2,Aliev3,Aliev4,Aliev5,Aliev6,Azizi:2012hs}. 

For years, the pseudoscalar $B_c$ meson was considered as the only type of  meson containing $b$ and $c$ quarks. But recently, other types of $B_c$ meson, such as scalar, vector, axial vector and tensor ones, although have not been discovered yet, are of interest and expected to be produced at LHCb in the future \cite{Godfrey:2004ya,Kolodziej:1995nv,Chang:1996jt,Cheung:1995ye,Cheung:1995ir,Gouz:2002kk,Chang:2003cr,Berezhnoy:2010wa,LHC}.
Some studies have been done on the mass and decay constants of the vector and axial-vector $B_c$ states \cite{Wang:2012kw} that can be used as input parameters to investigate different decay channels, which help us to determine their nature and structures.
Moreover, investigation of the decays of these mesons, provides windows for reliable determination of the CKM matrix elements, $V_{cb}$, and origin of the CP violation as well as looking for new physiscs effects.

The QCD sum rule is a profound theoretical tool to study many parameters of hadrons and their decay channels in the realms that ohter methods may fall into trouble \cite{SVZ}.
It has previousely applied successfully to determine many parameters of the hadrons and their interactions with other particles. In most of the cases, the predictions have been obtained to be in accord with the existing experimental data. For some of these studies see for instance the Refs. \cite{Azizi:2007jx,Agaev:2016dev,Aliev:2010ac,Aliev:2012ru,Aliev:2010uy,Agaev:2016mjb,Aliev:2009jt,Agaev:2016urs,Aliev:2012iv,Agaev:2017cfz,Sundu:2017xct,Azizi:2017izn,Azizi:2017ubq,Agaev:2017ywp,Azizi:2017bgs,Azizi:2017xyx,Agaev:2017lip,Agaev:2017jyt,Aliev:2016jnp,Aad:2014laa}.

In the present work, we comprehensively study the rare radiative decays of the vector and axial-vector mesons, $B^{(V,A)}_c$, to each of three charmed-strange $D_s$, $D^*_s$ and $D_{s1}$ mesons via light-cone QCD sum rule.
These decays can occure via two channels, weak annihilation (WA) and the electromagnetic penguin (EP) modes based on $b \rightarrow s \gamma$ at the quark level. 
We calculate the responsible form factors for all the channels under consideration and use them to estimate the relevant decay rates and branching ratios. As it was previously shown, for some rare radiative $B_c$ transitions, the EP parts are obtained to be at least three ordrs of magnitudes less than the WA channel contributions \cite{Azizi:2012hs} and can be ignored in the calculations. So we concentrate on just the WA modes of the considered decay channels.

For the photon, radiated from the final $D_s$, $D^*_s$ or $D_{s1}$ mesons,  we consider the bare loops as well as quark and quark-gluon condensates, and the propagation of the soft photon in the electromagnetic field. But for $B_{c}^{V(A)}$, as it contains just heavy quarks, the quark condensates and soft quark propagation are not included \cite{Aliev1,Colangelo:2000dp} and we only need to calculate contributions of the bare loop diagrams. 

The organization of the paper is as follows. In the next section we use the factorization hypothesis and Lorentz invariance to find the amplitudes of the considered decays in terms of the $B_{c}^{V(A)}$ and  $D_s$'s transition form factors. In section III we employ the QCD sum rule method to find the radiative form factors of  $B_{c}^{V(A)}$. In section IV the same procedure will be used to find  $D_s$'s form factors. Using some inputs and the sum rules obtained in the previous sections, in section V, we calculate the decay rates and branching ratios of the decays under considerations.

\section{TRANSITION AMPLITUDE FOR WEAK ANNIHILATION CHANNELS\label{Bound}}

\begin{figure}[tbp]
	\centering {\includegraphics[width=15cm]{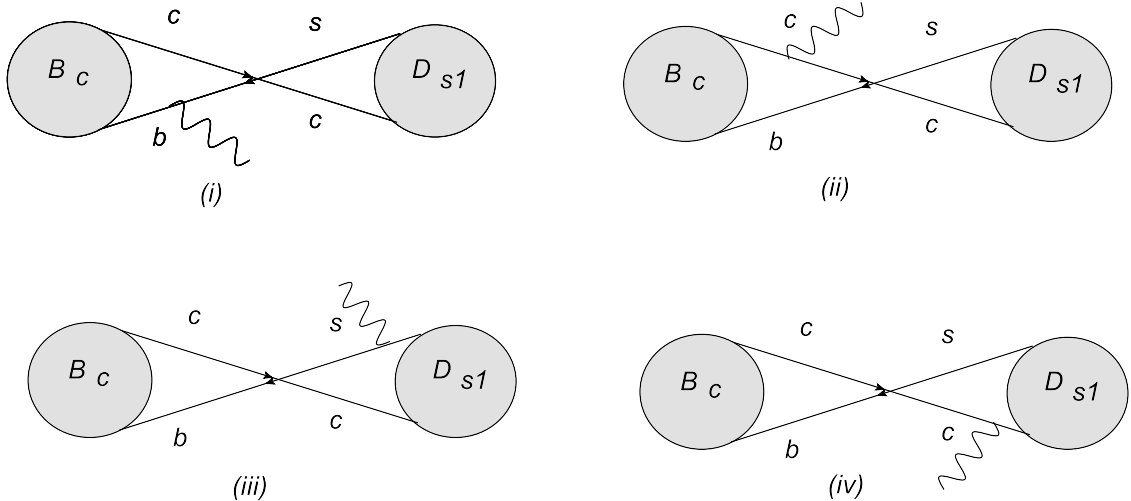}}
	\caption{Feynman diagrams for radiation of photon from either of initial and final states in quark level, representing  $B_{c} \to D_{s1} \gamma$, that are the same for other decays under considerations}
	\label{Fig1}
\end{figure}

In this section we concentrate on calculating the transition amplitudes in terms of the radiative form factors. The generic Feynman diagrams for $B_c$ to $D_{s1}$ transition, as an example, are shown in Fig. \ref{Fig1}.

The amplitude of these decays can be written as 
\begin{equation}
M(B_{c}\rightarrow D\gamma)=\frac{G_{F}}{\sqrt{2}}V_{cb}V_{cs}^{*}\langle D(p)\gamma(q)|(\overline{s}\Gamma_{\nu}c)(\overline{c}\Gamma^{\nu}b)|B_{c}(p+q)\rangle,  \label{weakamp}
\end{equation}
where $D(p)$ stands for any of $D_s$, $D^*_s$ or $D_{s1}$ mesons.

Now we use the factorization hypothesis to decompose the matrix elements in separate terms responsible for radiation of the photon from $B_c$ or $D_s$ mesons. For the case of pseudoscalar $D_s$ meson we set
\begin{eqnarray}
\langle D_s(p)\gamma(q)|(\overline{s}\Gamma_{\nu}c)(\overline{c}\Gamma^{\nu}b)|B_{c}(p+q)\rangle=-e\varepsilon^{\mu}p^{\nu}f_{D_{s}}T^{(B_{c})}_{\mu\nu}-ie\varepsilon^{\mu}\varepsilon_{(B_c)}^{\mu}m_{B_c}f_{B_{c}}T^{(D_{s})}_{\mu\nu}, \label{2}
\end{eqnarray}
while for the vector $D^*_s$ meson we obtain 
\begin{equation}
\langle D^*_s(p)\gamma(q)|(\overline{s}\Gamma_{\nu}c)(\overline{c}\Gamma^{\nu}b)|B_{c}(p+q)\rangle=-e\varepsilon^{\mu}\varepsilon_{(D^*_s)}^{\nu}m_{D^*_{s}}f_{D^*_{s}}T^{(B_{c})}_{\mu\nu}-ie\varepsilon^{\mu}\varepsilon_{(B_c)}^{\mu}m_{B_c}f_{B_{c}}T^{(D^*_{s})}_{\mu\nu}. \label{2'}
\end{equation}
The same is valid for the axial vector $D_{s1}$ by replacement, $D^*_s \rightarrow D_{s1}$, in the above formula.

In Eq.(\ref{2}), $T^{(B_c)}_{\mu\nu}$ in the first term stands for the emission of the photon from $B_c$ meson (diagrams (i) and (ii) in Fig. \ref{Fig1}) and the emission of the photon from $D_s$ meson is denoted by $T^{(D_s)}_{\mu\nu}$ in the second term (diagrams (iii) and (iv) in Fig. \ref{Fig1}). Also $f_{B_c}$ ($f_{D_s}$) is the $B_c$ ($D_s$) decay constant and $\varepsilon^\mu$ ($\varepsilon^{(B_c)\nu}$) is the polarization vector of the photon ($B_c$ meson).

The covariant amplitudes $T^{(B_c)}_{\mu\nu}$ and  $T^{(D_s)}_{\mu\nu}$ are defined by the following two-point correlation functions:
\begin{eqnarray}
T^{(B_c)}_{\mu\nu}(p,q)=i \int d^4 x e^{iq.x}\langle 0| T\lbrace j^{em}_{\mu}\bar{c}\Gamma_\nu b(0)\rbrace|B_c(p+q)\rangle , \label{3}
\end{eqnarray}
and
\begin{eqnarray}
T^{(D_s)}_{\mu\nu}(p,q)=i \int d^4 x e^{iq.x}\langle D_s(p)| T\lbrace j^{em}_{\mu}\bar{s}\Gamma_\nu c(0)\rbrace|0\rangle , \label{3}
\end{eqnarray}
where $j^{em}_{\mu}$ is the electromagnetic current.

Following the Ref. \cite{Wyler} we can write the $T^{(B_c)}_{\mu\nu}$ in the most general Lorentz covariant form as 
\begin{eqnarray}
T^{(B_c)}_{\mu\nu}(p,q) = ag_{\mu\nu} + bp_{\mu}p_{\nu} + cp_{\mu}q_{\nu} + dq_{\mu}p_{\nu} + eq_{\mu}q_{\nu}+ f\varepsilon_{\mu\nu\alpha\beta}p^{\alpha}q^{\beta} , \label{4}
\end{eqnarray}
where $a$, $b$, $c$, $d$, $e$ and $f$ are invariant amplitudes. 

Applying Ward identity for the electromagnetic current, alongside the fact that $q^2=0$ for the real photon and tranversity of the electromagnetic field, $\varepsilon.q=0$, we can write the first and the second terms in Eq.(\ref{2'}) as
\begin{eqnarray}
-e\varepsilon^{\mu}p^{\nu}f_{(D_s)}T^{(B_c)}_{\mu\nu} &&=-ef_{D_s}\Big[ \Big( (\varepsilon.\varepsilon^{(B_c)})(p.q) - (\varepsilon.p)(\varepsilon^{(B_c)}.q)\Big) i \dfrac{ F^{(B_c)}_{A}}{m^2_{B_c}} \notag \\ &&
- m_{B_c}f_{B_c}(\varepsilon.\varepsilon^{(B_c)}) + \varepsilon_{\nu\mu\lambda\sigma}\varepsilon^{\mu}\varepsilon^{(B_c)\nu}p^{\lambda}q^{\sigma}  \dfrac{ F^{(B_c)}_{V}}{m^2_{B_c}}\Big] , \label{5}
\end{eqnarray} 
and
\begin{eqnarray}
-ie\varepsilon^{\mu}\varepsilon^{(B_c)}m_{B_c}f_{B_c}T^{(D_s)}_{\mu\nu} &&=-iem_{B_c}f_{B_c}\Big[\Big((\varepsilon.\varepsilon^{(B_c)})(p.q)-(\varepsilon.p)(\varepsilon^{(B_c)}.q)\Big)i \dfrac{F^{(D_s)}_{A}}{m^2_{D_s}} \notag \\ &&
+ \dfrac{if_{D_s}}{(p.q)}(\varepsilon.p)(\varepsilon^{(B_c)}.p) + \varepsilon_{\nu\mu\lambda\sigma}\varepsilon^{\mu}\varepsilon^{(B_c)\nu}p^{\lambda}q^{\sigma} \dfrac{F^{(D_s)}_{V}}{m^2_{D_s}}\Big], \label{5'}
\end{eqnarray}
where $F^{(B_c)}_{V(A)}$ and $F^{(D_s)}_{V(A)}$ correspond to the parity coserving (parity violating) transition form factors.

Finally, combining Eqs. (\ref{5}), (\ref{5'}) and (\ref{weakamp}), and summing over the polarization vectors, we can get to the following result for the transition amplitude in $D_s$ case:
\begin{eqnarray}
M(B_c \rightarrow D_s \gamma) &&= -\dfrac{eG_F}{\sqrt{2}}V_{cb}V^*_{cs}\Big\{ f_{D_s}\Big[ \Big( (\varepsilon.\varepsilon^{(B_c)})(p.q) - (\varepsilon.p)(\varepsilon^{(B_c)}.q)\Big) i  \dfrac{ F^{(B_c)}_{A}}{m^2_{B_c}} - m_{B_c}f_{B_c}(\varepsilon.\varepsilon^{(B_c)}) \notag \\ &&
+ \varepsilon_{\nu\mu\lambda\sigma}\varepsilon^{\mu}\varepsilon^{(B_c)\nu}p^{\lambda}q^{\sigma}  \dfrac{F^{(B_c)}_{V}}{m^2_{B_c}} \Big] -im_{B_c}f_{B_c}\Big[ \Big( (\varepsilon.\varepsilon^{(B_c)})(p.q) - (\varepsilon.p)(\varepsilon^{(B_c)}.q)\Big)i \dfrac{F^{(D_s)}_{A}}{m^2_{D_s}} \notag \\ &&
+ \dfrac{i f_{D_s}}{(p.q)}(\varepsilon.p)(\varepsilon^{(B_c)}.p) + \varepsilon_{\nu\mu\lambda\sigma}\varepsilon^{\mu}\varepsilon^{(B_c)\nu}p^{\lambda}q^{\sigma}\dfrac{F^{(D_s)}_{V}}{m^2_{D_s}}\Big] \Big\}. \label{MBctoDs}
\end{eqnarray} 

Repeating the same procedure, and noticing that $D^*_s$ is a vector meson, we can write the transition amplitude for $B_c \rightarrow D^*_s \gamma$ as

\begin{eqnarray}
M(B_c \rightarrow D^*_s \gamma) &&=  \frac{e G_F V_{cb} V_{cs}}{\sqrt{2}}\Big\{ f_{B_c} m_{B_c} \Big[\Big((\varepsilon .\varepsilon ^{(B_c)})( q.\varepsilon^{(D^*_s)})-(q.\varepsilon^{(B_c)})( \varepsilon .\varepsilon^{(D^*_s)}) \Big) \frac{i F_A^{(D^*_s)} }{m_{D^*_s}^2}
\notag \\ && +f_{D^*_s} (\varepsilon.\varepsilon^{(D^*_s)})  + \varepsilon_{\mu\nu\lambda\sigma} q^{\mu} \varepsilon^{\nu}  \varepsilon ^{(B_c)\lambda} \varepsilon^{(D^*_s)\sigma} \frac{i F_A^{(D^*_s)} }{m_{D^*_s}^2} \Big]
\notag \\ && - f_{D^*_s} m_{D^*_s} \Big[\Big((\varepsilon .\varepsilon ^{(D^*_s)})( q.\varepsilon^{(B_c)})-(q.\varepsilon^{(D^*_s)})( \varepsilon .\varepsilon^{(B_c)}) \Big) \frac{i F_A^{(B_c)} }{m_{B_c}^2}
+f_{B_c} (\varepsilon.\varepsilon^{(B_c)}) \notag \\ && + \varepsilon_{\mu\nu\lambda\sigma} q^{\mu} \varepsilon^{\nu}  \varepsilon ^{(D^*_s)\lambda} \varepsilon^{(B_c)\sigma} \frac{i F_A^{(B_c)} }{m_{B_c}^2} \Big]\Big\},
\label{MBctoDss}
\end{eqnarray}
where can be used for the axial-vector meson $ D_{s1} $, as well.

In the next section we will calculate the transition form factors $F_{A(V)}^{(B_c)}$ of both the vector and axial-vector $B_c$ channels.

\section{LIGHT CONE QCD sum rule FOR THE VECTOR AND AXIAL VECTOR $B_c$ FORM FECRORS $F_{A(V)}^{(B_c)}$  \label{Bound}}

The general idea in QCD sum rule method is to calculate an appropriate correlation function both in phenomenological and theoretical representations and connecting them together via dispersion relation to find sum rules for physical quantities. For vector $B_c$ we write the correlation function as 

\begin{eqnarray}
\Pi^{B_c^{(V)}}_{\mu \nu}(p,q)=i\int d^4x e^{i Q.x} \langle \gamma(q)| T \{\bar{c}(x)\gamma_{\mu}(1-\gamma_5)b(x)\bar{b}(0)\gamma_{\nu}c(0)\}| 0\rangle,
\label{11}
\end{eqnarray} 
where $Q=p+q$ . To get to the hadronic (phenomenological) side, we insert a full set of hadronic $B_c$ states into Eq. (\ref{11}), and after integrating over $x$ we have:

\begin{eqnarray}
\Pi^{B_c^{(V)}}_{\mu \nu}(p,q)=\dfrac{\langle \gamma (q)| \bar{c} \gamma_{\mu} (1-\gamma_5) b | B_c(p+q) \rangle \langle B_c(p+q) | \bar{b} \gamma_{\nu} c | 0 \rangle}{m^2_{B_c}-(p+q)^2}.
\label{12}
\end{eqnarray}
The second bracket in Eq.(\ref{12}) can be written by means of

\begin{eqnarray}
\langle B_c(p+q) | \bar{b} \gamma_{\nu} c | 0 \rangle = f_{B_c} m_{B_c} \varepsilon_{\nu}^{(B_c)}.
\label{13}
\end{eqnarray}
As the first bracket in Eq. \ref{12} contains both vector ($\gamma_{\mu}$) and axial-vector ($\gamma_{\mu} \gamma_5$) parts, considering the parity properties of vector $B_c$ meson ($J^P=1^-$), Lorentz and gauge invariance, we can write it in two terms as

\begin{eqnarray}
\langle \gamma (q)| \bar{c} \gamma_{\mu} (1-\gamma_5) b | B_c(p+q) \rangle = e \Big\{ i \varepsilon_{\mu\alpha\beta\sigma} \varepsilon^{\alpha} \varepsilon^{(B_c)} q^{\sigma} \frac{F^{(B_c)}_{V}(Q^2)}{m^2_{B_c}}  \notag \\  
+[\varepsilon_{\mu}(\varepsilon^{(B_c)}.q) - q_{\mu} (\varepsilon . \varepsilon^{(B_c)})] \frac{F^{(B_c)}_{A}(Q^2)}{m^2_{B_c}} \Big\}.
\label{14}
\end{eqnarray}

By substituting Eqs. (\ref{13}) and (\ref{14}) into Eq. (\ref{12}) we can write the hadronic side as 

\begin{eqnarray}
\Pi_{\mu \nu}^{B_c^{(V)}}(p,q) = \frac{e f_{B_c} m_{B_c}}{m^2_{B_c}-Q^2} \Big\{ i \varepsilon_{\mu\nu\alpha\sigma} \varepsilon^{\alpha} q^{\sigma} \frac{F^{(B_c)}_{V}(Q^2)}{m^2_{B_c}} + [q_{\mu} \varepsilon_{\nu} - \varepsilon_{\mu} q_{\nu}] \frac{F^{(B_c)}_{A}(Q^2)}{m^2_{B_c}} \Big\}.
\label{15}
\end{eqnarray}

Now, to calculate the theoretical side, we write the correlation function in terms of two structures given in (\ref{15}) as follows:

\begin{eqnarray}
\Pi_{\mu\nu}^{B_c^{(V)}}(p,q) =  i \varepsilon_{\mu\nu\alpha\sigma} \varepsilon^{\alpha} q^{\sigma} \Pi_{1} +  [q_{\mu} \varepsilon_{\nu} - \varepsilon_{\mu} q_{\nu}] \Pi_{2},
\label{16}
\end{eqnarray}
where each functions $\Pi_{1}$ and $\Pi_{2}$ have perturbative and non-perturbative contributions as follows:

\begin{eqnarray}
\Pi_i=\Pi^{\text{pert.}}_i+\Pi^{\text{non-pert.}}_i
\label{17}
\end{eqnarray}

The perturbative part contains bare loops and non-perturbative part gets its contribution from quark condensates, quark-gluon condensates, and soft photon in electromagnetic field. But as $B_c$ contains two heavy quarks, these non-perturbative parts have no contribution in it \cite{Colangelo:2000dp,Aliev1}. 
So, for both vector and axial-vector $B_c$ mesons, we just need to calculate the bare loop contribution.

To calculate the perturbative parts, we consider Fig.2 (a) and Fig.2 (b), when the photon is radiated both from $b-$ or $c-$quark. These structures can be related to the spectral density using the Cutkosky method (dispersion relation) as

\begin{eqnarray}
\Pi_{1(2)}^{\text{pert.}}(p,q) = \int ds \dfrac{\rho_{1(2)}(s,p^2)}{s-p^2} + \text{subtraction  terms},
\label{18}
\end{eqnarray}
where $\rho_{i}$'s are the spectral densities. 

To calculate  $\rho_{i}$'s, we write the correlation function using Feynman rules for Fig.2 (a) as:

\begin{eqnarray}
\Pi^{B_c^{(V)}}_{\mu \nu (a)} = e N_c Q_s \int \dfrac{d^4 k}{(2 \pi)^4} \Big\{ \text{Tr} \Big[ \dfrac{i(\not\!k + m_c)}{k^2 - m_c^2} \gamma_{\mu} (1-\gamma_5)
\notag \\ \times \dfrac{i(\not\!p +\not\!k + m_b)}{(p+k)^2 - m_b^2}  \not\!\varepsilon \dfrac{i(\not\!Q + \not\!k + m_b)}{(Q+k)^2 - m_b^2} \gamma_{\nu} \Big] \Big\}.
\label{19}
\end{eqnarray}

Now, using Feynman parametrization we can write the coefficients of the structures $i\varepsilon_{\mu\nu\alpha\beta}q^{\alpha}p^{\beta}$ and $[q_{\mu}\varepsilon_{\nu}-q_{\nu}\varepsilon_{\mu}]$ as 

\begin{eqnarray}
\Pi_{1(a)}^{(\text{pert.})} = -\dfrac{e N_c Q_b}{4 \pi^2} \Big\{ \int_{0}^{1} dy \int_{0}^{1}  dx   x [m_b (m_c + m_b x \bar{y}) + x \bar{x} (x \bar{y} + 
y) p.p 
\notag \\  + 2 x^2\bar{x} \bar{y}^2 p.q] \int_{0}^{\infty} d\alpha e^{- \alpha \Delta} \Big\},
\label{20}
\end{eqnarray}

and

\begin{eqnarray}
\Pi_{2(a)}^{(\text{pert.})} = \dfrac{e N_c Q_b}{4 \pi^2} \Big\{  \int_{0}^{1} dy \int_{0}^{1}  dx x [m_b (m_c - m_b x \bar{y}) +  x \bar{x} (2 - y - x \bar{y}) p.p 
\notag \\ + 2  x \bar{x} \bar{y} (1 - x \bar{y}) p.q
]  \int_{0}^{\infty} d\alpha e^{- \alpha \Delta} \Big\},
\label{21}
\end{eqnarray}
where $\bar{x}( \bar{y} ) =1-x(y) $ and $\Delta = m_c^2 \bar{x} + m_b^2 x - x \bar{x} y  p.p - x \bar{x} \bar{y} Q.Q$, and we have used the Schwinger parametrization:

\begin{eqnarray}
\frac{1}{\Delta^n} = \int_{0}^{\infty} d \alpha e^{- \alpha \Delta}.
\label{22}
\end{eqnarray}

Now we apply a double Borel transformation, $\hat{B}_{Q^2}(M_1^2) \hat{B}_{p^2}(M_2^2)$ on $\Pi_{i}^{\text{pert.}}$ that transforms $Q^2\rightarrow M_1^2$ and $p^2\rightarrow M_2^2$ which yields

\begin{eqnarray}
\hat{\Pi}_{1(a)}^{\text{pert.}} =&& \frac{e  N_c Q_b}{4 \pi ^2}\frac{\sigma _1 \sigma _2}{\left(\sigma _1+\sigma _2\right){}^2} \int_{0}^{1} dx \frac{1}{\bar{x}} e^{-\frac{\left(m_c^2 \bar{x}+m_b^2 x\right)\left(\sigma _1+\sigma _2\right)}{x \bar{x}}} 
\notag \\  \times && \Big[m_b^2 x \left(\sigma _1 x + \sigma _2\right)+m_c^2 \bar{x} \left(\sigma _1 x + \sigma _2\right)-x \bar{x} \frac{\sigma _1-\sigma _2 }{\sigma _1+\sigma _2}\Big],
\label{23}
\end{eqnarray}
and 
\begin{eqnarray}
\hat{\Pi}_{2(a)}^{\text{pert.}} =&& \dfrac{e N_c Q_b}{4 \pi^2} \dfrac{ \sigma_1 \sigma_2 }{\left(\sigma _1+\sigma _2\right){}^2} \int_{0}^{1} dx \frac{1}{\bar{x}} e^{-\frac{(m_c^2 \bar{x} + m_b^2 x) (\sigma_1 + \sigma_2)}{x \bar{x}}}
\notag \\ \times && \Big[m_c^2 \bar{x} \left(\sigma _1 (2-x)+\sigma _2\right) + m_b^2 x \left(\sigma _1 (2-x)+\sigma _2\right)+ x \bar{x}\frac{ 2\sigma _1 }{\sigma _1+\sigma _2} \Big],
\label{24}
\end{eqnarray}
where $\sigma_{1(2)}=1/M_{1(2)}^2$. 
To perform the Borel transformation we have used the following identity:

\begin{eqnarray}
\hat{B}(M^2) e^{- \alpha p^2} = \delta (1 - \alpha M^2).
\label{25}
\end{eqnarray}

To enhance the contribution of the ground satates, we apply a second double Borel transformation on $\hat{\Pi}_{i(a)}^{\text{pert.}}$ that transforms $\sigma_{1}$ and $\sigma_{2}$ to the new variables $s$ and $t$ respectively as follows:

\begin{eqnarray}
\hat{B}(\frac{1}{s} , \sigma_{1}) \hat{B}(\frac{1}{t} , \sigma_{2}) e^{- \alpha (\sigma_{1} + \sigma_{2})} = \delta (1 - \frac{\alpha}{s}) \delta (1 - \frac{\alpha}{t}) ,
\label{26}
\end{eqnarray}
to get

\begin{eqnarray}
\varrho_{i} (s,t) = \frac{1}{s t} \hat{B}(\frac{1}{s} , \sigma_{1}) \hat{B}(\frac{1}{t} , \sigma_{2}) \dfrac{\hat{\Pi}^{\text{pert.}}_{i}}{\sigma_{1} \sigma_{2}}.
\label{27}
\end{eqnarray}
Substituting $\varrho_{i}(s,t)$ in the relation 

\begin{eqnarray}
\rho_{i}(s,p^2) = \int dt \dfrac{\varrho_{i} (s,t)}{t - p^2},
\label{28}
\end{eqnarray}
and after lengthy calculations we get for the spectral densities:

\begin{eqnarray}
\rho_{1(a)}^{B_c^{(V)}} (s , p^2) =&& \frac{e N_c Q_b}{4 \pi^2} \dfrac{1}{(s-p^2)^3} \int_{x_0}^{x_1} \frac{1}{x^2\bar{x}^3} \Big\{(x+1) \left(x m_b^2+\bar{x} m_c^2\right) \Big(x \left(m_b^2-p^2 \bar{x}\right)+\bar{x} m_c^2\Big)
\notag \\ \times && \Big(x \left(m_b^2-s \bar{x}\right)+\bar{x} m_c^2\Big)\Big\},
\label{29}
\end{eqnarray} 
and
\begin{eqnarray}
\rho_{2(a)}^{B_c^{(V)}} (s , p^2) =&& \frac{e N_c Q_b}{4 \pi^2} \dfrac{1}{(s-p^2)^3} \int_{x_0}^{x_1} \frac{1}{x^2 \bar{x}^3} \Big\{ (3-x) \left(x m_b^2+\bar{x} m_c^2\right) \Big(x m_b^2+\bar{x} \left(m_c^2-p^2 x\right)\Big)
\notag \\ \times && \Big(x m_b^2+\bar{x} \left(m_c^2-s x\right)\Big) \Big\}.
\label{30}
\end{eqnarray} 
The integral boundaries $x_0$ and $x_1$ need to satisfy the following inequality:
\begin{eqnarray}
s x \bar{x} - (m_c^2 \bar{x} + m_b^2 x) \geq 0
\label{31}
\end{eqnarray} 
coming from the constraint imposed by the integral over the $\delta$-function.

To calculate the contribution from the Fig.2 (b) we just need to interchange the $b$- and $c$- quark parameters in Eqs.(\ref{29}) and (\ref{30}). Finally by adding the contributions of Fig.2 (a) and 2 (b), the corresponding two selected structures are

\begin{eqnarray}
\rho_1^{B_c^{(V)}}(s , p^2) &&= \dfrac{e N_c s^2}{8 \pi^2 (s-p^2)^3} \Bigg\{ Q_c \Bigg[\lambda  \Big[p^2 \Big(\alpha  (\alpha +9)-(2 \alpha +3) \beta +\beta ^2\Big)+2 \alpha  s (3 \alpha -7 \beta +1)\Big]
\notag \\ && -2 \alpha  \Big[p^2 (3 \alpha -4 \beta +2)+s \Big(\alpha  (\alpha +3)-\beta(3 \alpha +4) +6
\beta ^2\Big)\Big] \text{Ln}\left(\frac{1+\alpha -\beta +\lambda }{1+\alpha -\beta -\lambda}\right) 
\notag \\ && +2 \beta ^2 \Big(p^2-s (3 \alpha +\beta -1)\Big) \text{Ln}\left(\frac{1-\alpha +\beta +\lambda }{1-\alpha +\beta -\lambda }\right)\Bigg] \notag \\ && +Q_b \Bigg[\lambda  \Big[p^2 \Big(\beta  (\beta +9)-(2 \beta +3) \alpha+\alpha ^2\Big)+2 \beta  s (3 \beta -7 \alpha +1)\Big]
\notag \\ && -2 \beta  \Big[p^2 (3 \beta -4 \alpha +2)+s \Big(\beta  (\beta +3)-\alpha(3 \beta +4) +6
\alpha ^2\Big)\Big] \text{Ln}\left(\frac{1+\beta -\alpha +\lambda }{1+\beta -\alpha -\lambda}\right) 
\notag \\ && +2 \alpha ^2 \Big(p^2-s (3 \beta +\alpha -1)\Big) \text{Ln}\left(\frac{1-\beta +\alpha +\lambda }{1-\beta +\alpha -\lambda }\right)\Bigg] \Bigg\},
\label{32}
\end{eqnarray}     

and 
\begin{eqnarray}
\rho_2^{B_c^{(V)}}(s , p^2) &&= \dfrac{e N_c s^2}{8 \pi^2 (s-p^2)^3} \Bigg\{  Q_b \Bigg[ \lambda  \Big[p^2 \Big( \alpha (\alpha -7)-2 \alpha  \beta  +\beta ^2+5 \beta \Big)+2 s \Big(  \alpha(\alpha -1)+ 3\alpha  \beta +4 \beta ^2\Big)\Big]
\notag \\ &&+ 2 \alpha \Big[p^2 (\alpha -4 \beta +2)+s\Big( \alpha(1-\alpha ) +3 \alpha  \beta  +2 \beta  (3 \beta -2) \Big)\Big] \text{Ln}\left(\frac{1+\alpha -\beta +\lambda}{1+\alpha -\beta -\lambda }\right) \notag \\ && -2 \beta ^2 \Big(3 p^2- s (9 \alpha -\beta -3)\Big) \text{Ln} \left(\frac{1-\alpha +\beta +\lambda}{1-\alpha +\beta -\lambda}\right)  \Bigg]
\notag \\ && + Q_c \Bigg[\lambda  \Big[p^2 \Big( \beta (\beta -7)-2 \beta  \alpha  +\alpha ^2+5 \alpha \Big)+2 s \Big(  \beta(\beta -1)+ 3\beta  \alpha +4 \alpha ^2\Big)\Big]
\notag \\ &&+ 2 \beta \Big[p^2 (\beta -4 \alpha +2)+s\Big( \beta(1-\beta ) +3 \beta  \alpha  +2 \alpha  (3 \alpha -2) \Big)\Big] \text{Ln}\left(\frac{1+\beta -\alpha +\lambda}{1+\beta -\alpha -\lambda }\right) \notag \\ && -2 \alpha ^2 \Big(3 p^2- s (9 \beta -\alpha -3)\Big) \text{Ln} \left(\frac{1-\beta +\alpha +\lambda}{1-\beta +\alpha -\lambda}\right) \Bigg] \Bigg\},
\label{33}
\end{eqnarray} 
where $\alpha=\frac{m_b^2}{s}$ and $\beta=\frac{m_c^2}{s}$ and $\lambda=\sqrt{1+\alpha^2+\beta^2-2\alpha -2\beta - 2\alpha \beta}$.

To calculate these structures for the axial $B_c$ meson we write the corresponding correlation function as
\begin{eqnarray}
\Pi_{\mu \nu}^{B_c^{(A)}} &=& e N_c Q_s \int \dfrac{d^4 k}{(2 \pi)^4} \Big\{ \text{Tr} \Big[ \dfrac{i(\not\!k + m_c)}{k^2 - m_c^2} \gamma_{\mu} (1-\gamma_5)
\notag \\ && \times \dfrac{i(\not\!p +\not\!k + m_b)}{(p+k)^2 - m_b^2}  \not\!\varepsilon \dfrac{i(\not\!Q + \not\!k + m_b)}{(Q+k)^2 - m_b^2} \gamma_{\nu}\gamma_5 \Big] \Big\},
\label{34}
\end{eqnarray}
where $(A)$ stands for the axial-vector $B_c$ meson.

Following similar procedure done for vector $B_c$, we can calculate the spectral densities as

\begin{eqnarray}
\rho^{B_c^{(A)}}_1 (s , p^2) &&= \dfrac{e N_c s^2}{8 \pi^2 (s - p^2)^3} \Bigg\{ Q_b \Bigg[\lambda  \Big[p^2 \Big(\alpha(\alpha -7) -2 \alpha \beta +\beta \left( 1+5 \beta\right) \Big) -2s \Big(\alpha (1-\alpha)-3 \alpha  \beta -4 \beta ^2\Big)\Big] 
\notag \\ && +2 \alpha \Big[p^2 (\alpha -4 \beta +2)+s \Big(\alpha (1-\alpha ) +3 \alpha  \beta +2 \beta  (3
\beta -2)\Big)\Big] \text{Ln}\left(\frac{1+\alpha -\beta +\lambda}{1+\alpha -\beta -\lambda}\right) 
\notag \\ && -2 \beta ^2  \Big(3 p^2+s (-9 \alpha +\beta +3)\Big) \text{Ln} \left(\frac{1-\alpha +\beta +\lambda }{1-\alpha +\beta -\lambda }\right) \Bigg] + \notag \\ &&
Q_c \Bigg[  \lambda  \Big[p^2 \Big(\beta(\beta -7) -2 \beta \alpha +\alpha \left( 1+5 \alpha\right) \Big) -2s \Big(\beta (1-\beta)-3 \beta  \alpha -4 \alpha ^2\Big)\Big] 
\notag \\ && +2 \beta \Big[p^2 (\beta -4 \alpha +2)+s \Big(\beta (1-\beta ) +3 \beta  \alpha +2 \alpha  (3
\alpha -2)\Big)\Big] \text{Ln}\left(\frac{1+\beta -\alpha +\lambda}{1+\beta -\alpha -\lambda}\right) 
\notag \\ && -2 \alpha ^2  \Big(3 p^2+s (-9 \beta +\alpha +3)\Big) \text{Ln} \left(\frac{1-\beta +\alpha +\lambda }{1-\beta +\alpha -\lambda }\right) \Bigg] \Bigg\},
\label{35}
\end{eqnarray} 
and

\begin{eqnarray}
\rho^{B_c^{(A)}}_2 (s , p^2) &&= \dfrac{e N_c s^2}{8 \pi^2 (s - p^2)^3}  \Bigg\{ Qb \Bigg[ \lambda  \Big[ p^2 \Big(\alpha(7-\alpha )+2 \alpha  \beta  -\beta (5-\beta) \Big)+2 s \Big(\alpha(1-\alpha )-3 \alpha  \beta  -4 \beta ^2\Big)\Big]
\notag \\ && -2 \alpha \Big[ p^2 (\alpha -4 \beta +2)+s \Big(  \alpha (1-\alpha ) +3 \alpha  \beta -2  \beta (2-3 \beta
) \Big)\Big] \text{Ln}\left(\frac{1+\alpha -\beta +\lambda}{1+\alpha -\beta -\lambda}\right)
\notag \\ && 2 \beta ^2 \Big(3 p^2 -s (9 \alpha -\beta -3)\Big)  \text{Ln}\left(\frac{1-\alpha +\beta +\lambda}{1-\alpha +\beta -\lambda}\right)  \Bigg]
\notag \\ && + Qc \Bigg[ \lambda  \Big[ p^2 \Big(\beta(7-\beta )+2 \beta  \alpha  -\alpha (5-\alpha) \Big)+2 s \Big(\beta(1-\beta )-3 \beta  \alpha  -4 \alpha ^2\Big)\Big]
\notag \\ && -2 \beta \Big[ p^2 (\beta -4 \alpha +2)+s \Big(  \beta (1-\beta ) +3 \beta  \alpha -2  \alpha (2-3 \alpha
) \Big)\Big] \text{Ln}\left(\frac{1+\beta -\alpha +\lambda}{1+\beta -\alpha -\lambda}\right)
\notag \\ && 2 \alpha ^2 \Big(3 p^2 -s (9 \beta -\alpha -3)\Big)  \text{Ln}\left(\frac{1-\beta +\alpha +\lambda}{1-\beta +\alpha -\lambda}\right)  \Bigg]\Bigg\}.
\label{36}
\end{eqnarray}

Now, we are ready to find the QCD sum rule for $B_c$ form factors. By matching the selected structures from both QCD and hadronic sides and performing the Borel transformation with respect to $Q^2$ $(Q^2\rightarrow M_B^2)$, and also using quark-hadron duality we get the following generic result for $B_c$ form factors:

\begin{eqnarray}
F^{B_c^{(V/A)}}_{V,A} (p^2) =\frac{f_{B_c}}{e~ m_{B_c}} e^{m_{B_c}/M^2} \widehat{B}_{Q^2} \Big[ \int_{(m_b + m_c)^2}^{s_0} ds \dfrac{\rho^{B_c^{(V/A)}}_{1,2}(s,p^2)}{s-Q^2} \Big],
\label{37}
\end{eqnarray} 
where $s_0$ is the continuum threshold and the subscript $V(A)$ on the left hand side of Eq.(\ref{37}) corrsponds to the subscript $1(2)$ on the right hand side, and the form factors have to be evaluated at $p^2=m^2_{D_s}$.

The following standard rule for the Borel transformation has been used to get to the result of Eq.(\ref{37}):

\begin{eqnarray}
\hat{B}_{M_B^2}\Big(\dfrac{1}{(p^2-s)^n}\Big)=(-1)^n \dfrac{e^{-s/M_B^2}}{\Gamma(n)(M_B^2)^n}.
\label{38}
\end{eqnarray}

\section{LIGHT CONE QCD sum rule FOR THE FORM FACTORS $F_{V(A)}^{(D_s)}$, $F_{V(A)}^{(D^*_s)}$ AND $F_{V(A)}^{(D_{S1})}$\label{Bound}}

\begin{figure}[tbp]
	\centering {\includegraphics[width=15cm]{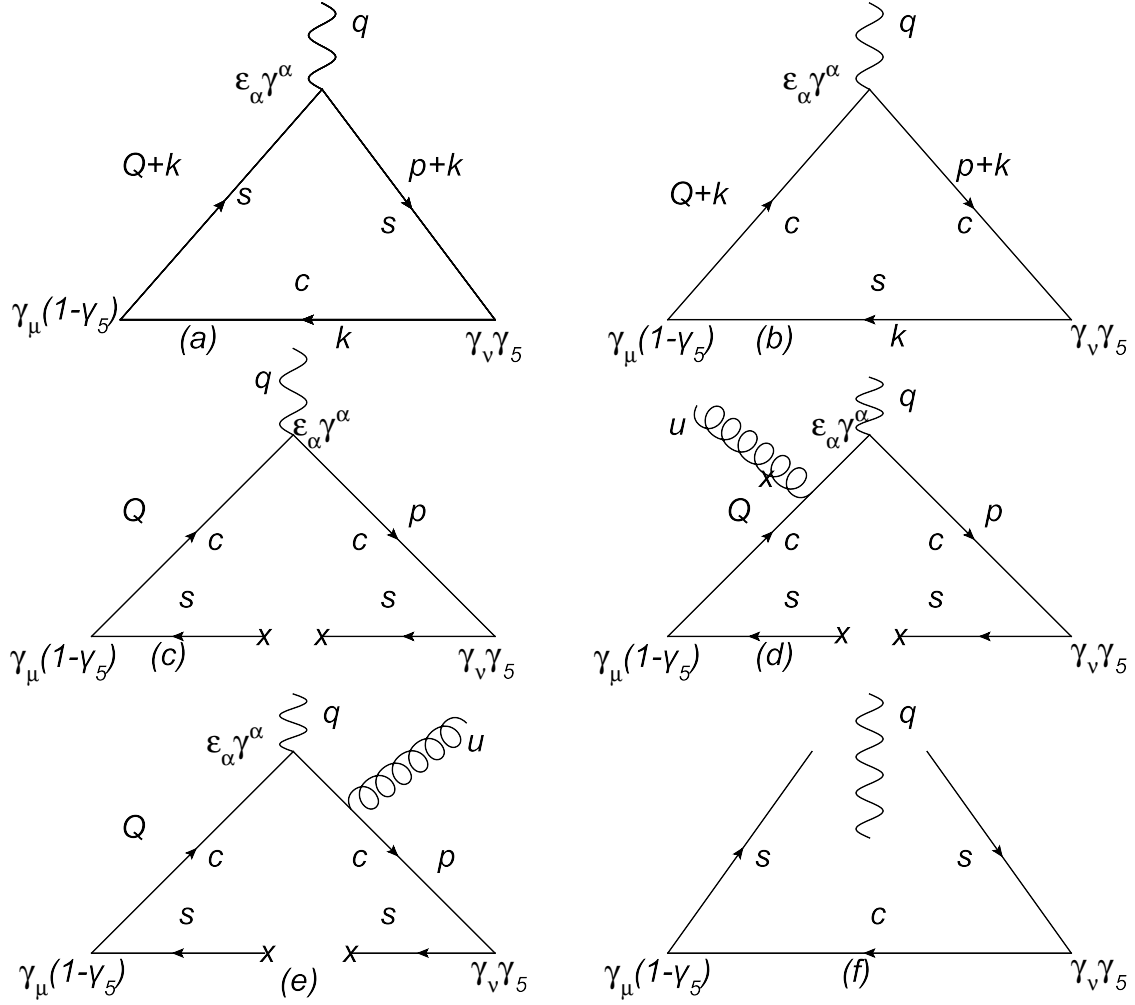}}
	\caption{Feynnman diagrams for perturbative contribitions (bare-loop) [(a), (b)], and non-perturbative contributions, including quark condensates (c), quark-gluon condensates [(d), (c)], and propagation of soft photon in electromagnetic field (f), for the final state $D_{s1} $ meson, that are the same as that of $D_s$ or $D^*_s$.}
	\label{F2}
\end{figure}

As $D_s$, $D^*_s$ and $D_{s1}$ mesons contain one light quark (s-quark), the non-perturbative parts (like quark condensates, quark-gluon condensates and soft photon in the electromagnetic field) contribute in calculating the relevant form factors, as shown for instance in Fig.\ref{F2} for $D_{s1}$ meson. But the line for calculating the perturbative part is the same as what is done in the last section for $B_c$ mesons. So we just write the final expressions for the spectral densities of these mesons. We should note that as $D_s$ is a pseudoscalar meson, its invariant structures are to some extent different from $D^*_s$ and $D_{s1}$ that are vector and axial-vector mesons respectively.

To find the relevant invariant structures for $D_s$ meson, we write the corresponding correlation function and insert the full set of hadronic states of $D_s$ meson and we get to

\begin{eqnarray}
\Pi_{\mu}^{(D_s)} &&= \dfrac{i e f_{D_s} m_{D_s}}{m_c + m_s} \dfrac{1}{m^{2}_{D_s} - p^2} \Big\{ i \varepsilon_{\mu\alpha\beta\sigma} \varepsilon^{\alpha} p^{\beta} q^{\alpha} \frac{F^{(D_s)}_{V}(Q^2)}{m_{D_s}^2}
\notag \\ && + [ \varepsilon_{\mu}(p.q) - q_{\mu} (\varepsilon . p) ] \frac{F^{(D_s)}_{A}(Q^2)}{m_{D_s}^2} \Big\}.
\label{39}
\end{eqnarray}

So for the QCD part, the correlation function in terms of the Lorentz invariant structures can be written as:

\begin{eqnarray}
\Pi_{\mu}^{(D_s)} = i \varepsilon_{\mu\alpha\beta\sigma} \varepsilon^{\alpha} p^{\beta} q^{\alpha} \Pi^{(D_s)}_{1} + [ \varepsilon_{\mu}(p.q) - q_{\mu} (\varepsilon . p) ] \Pi^{(D_s)}_{2}.
\label{40}
\end{eqnarray}

Following the same line as we did for $B_c$ mesons, the spectral density corresponding to the perturbative part of $\Pi_i$ structures would be as follows:

\begin{eqnarray}
\rho_1^{(D_s)} (t,Q^2) &&= \frac{e N_c}{4 \pi ^2 (t-Q^2)} \Bigg\{Q_s \left[\lambda  \left(m_c-m_s\right)+m_s \text{Ln}\left(\frac{1+\alpha -\beta +\lambda}{1+\alpha -\beta -\lambda}\right)\right]
\notag \\ &&+Q_c \left[\lambda  \left(m_s-m_c\right)+m_c \text{Ln}\left(\frac{1+\beta -\alpha +\lambda}{1+\beta -\alpha -\lambda }\right)\right]\Bigg\},
\label{41}
\end{eqnarray}
and 
\begin{eqnarray}
\rho_2^{(D_s)} (t,Q^2) &&= \frac{e N_c}{4 \pi ^2 (t-Q^2)^2} \Bigg\{ Q_s \bigg[ \lambda  \Big(m_c \left( (\alpha -\beta )t -Q^2\right)-m_s \left(t-Q^2\right)\Big) 
\notag \\ && + m_s (2
m_c m_s+t-Q^2) \text{Ln} \left(\frac{1+\alpha -\beta +\lambda}{1+\alpha -\beta -\lambda}\right)\bigg]
\notag \\ &&  +Q_c \bigg[ \lambda  \Big(m_s \left( (\beta -\alpha )t -Q^2\right)-m_c \left(t-Q^2\right)\Big) 
\notag \\ && + m_c (2
m_s m_c+t-Q^2) \text{Ln} \left(\frac{1+\beta -\alpha +\lambda}{1+\beta -\alpha -\lambda}\right)\bigg] \Bigg\}.
\label{42}
\end{eqnarray}

As shown in Fig.(\ref{F2}), the non-perturbative parts of $\Pi_i$, containing quark condensate and quark-gluon condensate contribute. 

After a lengthy but straightforward calculations, the total non-perturbative contributions corresponding to the relevant diagrams of Fig.(\ref{F2}) are as follows:

\begin{eqnarray}
\Pi_{1(c+d+e)}^{(D_s)} &&= \frac{e Q_c \langle\bar{s}s\rangle}{12 r^6 R^6} \Big[m_0^2 \Big(6 m_c^2 \left(r^4+r^2 R^2+R^4\right)-5 r^4 R^2+9 r^2 R^4\Big)-12 r^4 m_c^2 m_s^2 
\notag \\ && +6 r^2 R^2 m_s \Big(r^2 \left(m_c+2 m_s\right)-2 m_c^2 m_s\Big)  \notag \\ && -6 R^4\Big(r^2 m_s \left(m_c+2 m_s\right) + 2 m_c^2 m_s^2 + 2 r^4\Big)\Big],
\label{43}
\end{eqnarray}

\begin{eqnarray}
\Pi_{2(c+d+e)}^{(D_s)} &&= \frac{e Q_c \langle\bar{s}s\rangle}{6 r^6 R^6}
\Big[  m_0^2 \Big(m_c^2 \left(r^4+3 R^4\right)-3 r^2 \left(r^4+r^2 R^2+R^4\right)\Big)+3 r^4 R^2 m_s \left(2 m_s-m_c\right) 
\notag \\ && +3 R^4 \Big(r^2 m_s \left(m_c+m_s\right)-2 m_c^2
m_s^2-2 r^4\Big)+9 r^6 m_s^2 \Big],
\label{44}
\end{eqnarray}
where $R^2=Q^2 - m_c^2 $ and $r^2=p^2 - m_c^2$.

The final non-perturbative contribution to the emission of the photon from final state meson  ($D_s$, $D^*_s$ or $D_{s1}$) is the soft photon in the electromagnetic field as shown in Fig. \ref{F2}(f).

To calculate the invariant structures corresponding to the Feynman diagram of Fig. \ref{F2}(f), we write the vacuume to photon correlation function as:

\begin{eqnarray}
\Pi_{\mu(f)}^{(D_s)}(p,q)=i\int d^{4}x e^{-iQx} \langle\gamma(q)|T\{\overline{s}(0)\gamma_{5}c(0)\overline{c}(x)\gamma_{\mu}(1-\gamma_{5})s(x)\}|0\rangle,
\label{45}
\end{eqnarray}
and after contracting the c-quark lines we get

\begin{eqnarray}
\Pi_{\mu(f)}^{(D_s)}(p,q) = i^2\int d^{4}x \frac{d^{4}k}{(2\pi)^4}\frac{e^{-i(Q-k)x}}{m_{c}^2-k^2} \langle\gamma(q)|\overline{s}\gamma_{5}(\not\!k+m_{c})\gamma_{\mu}(1-\gamma_{5})s|0\rangle .
\label{46}
\end{eqnarray}

To write the correlation function in terms of photon distribution amplitudes (PDAs) we use the following $\gamma$-matrix identities:

\begin{eqnarray}
\gamma_{\mu}\gamma_{\nu} &=& g_{\mu\nu}+i\sigma_{\mu\nu}, \notag \\
\gamma_{\mu}\gamma_{\nu}\gamma_{5} &=& g_{\mu\nu}\gamma_{5}-\frac{i}{2}\varepsilon_{\mu\nu\alpha\beta}\sigma_{\alpha\beta} , \\
\gamma_{\mu}\gamma_{\alpha}\gamma_{\nu} &=& g_{\mu\alpha}\gamma_{\nu}+g_{\nu\alpha}\gamma_{\mu}-g_{\mu\nu}\gamma_{\alpha}+i\varepsilon_{\mu\nu\alpha\lambda}\gamma_{\lambda}\gamma_{5},\notag
\label{47}
\end{eqnarray}

The required PDAs are defined as \cite{DA1,DA2}
\begin{eqnarray}
\langle\gamma(q)|\overline{s}\gamma_{\nu}s|0\rangle &=& -\frac{Q_{s}}{2}f_{3\gamma}\int_{0}^{1}du \overline{\psi}^{(V)}(u)x^{\theta}F_{\theta\nu}(u x), \notag \\
\langle\gamma(q)|\overline{s}\gamma_{\alpha}\gamma_{5}s|0\rangle &=& -\frac{iQ_{s}}{4}f_{3\gamma}\int_{0}^{1}du \overline{\psi}^{(A)}(u)x^{\theta}\widetilde{F}_{\theta\alpha}(u x), \\
\langle\gamma(q)|\overline{s}\sigma_{\alpha\beta}s|0\rangle &=& Q_{s}\langle\overline{s}s\rangle\int_{0}^{1}du\phi(u)F_{\alpha\beta}(ux)\notag \\ &&\
+\frac{Q_{s}\langle\overline{s}s\rangle}{16}\int_{0}^{1}du x^{2}\textbf{A}(u)F_{\alpha\beta}(ux)\notag \\ &&\
+\frac{Q_{s}\langle\overline{s}s\rangle}{8}\int_{0}^{1}du \textbf{B}(u)x^{\rho}(x_{\beta}F_{\alpha\rho}(ux)-x_{\alpha}F_{\beta\rho}),\notag
\label{50}
\end{eqnarray}
where the functions $A(u)$, $B(u)$ $\overline{\psi}^{(V)}(u)$, and $\overline{\psi}^{(A)}(u)$ are as follows:

\begin{eqnarray}
\overline{\psi}^{(V)}(u) &=& -20u(1-u)(2u-1)+\frac{15}{16}(\omega_{\gamma}^{A}-3\omega_{\gamma}^{V})u(1-u)(2u-1)(7(2u-1)^2-3),\notag \\
\overline{\psi}^{(A)}(u) &=& (1-(2u-1)^2)(5(2u-1)^2-1)\frac{5}{2}(1+\frac{19}{16}\omega_{\gamma}^{V}-\frac{3}{16}\omega_{\gamma}^{A}),\notag \\
\textbf{A}(u) &=& 40u(1-u)(3k-k^{+}+1)+8(\xi_{2}^{+}-3\xi_{2})\notag \\ &&\
\times[u(1-u)(2+13u(1-u))+2u^3(10-15u+16u^2)\ln u \\ &&\
+2(1-u)^3(10-15(1-u)+6(1-u^2))\ln(1-u)],\notag \\
\textbf{B}(u) &=& 40\int_{0}^{u}d\alpha(4-\alpha)(1+3k^{+})[-\frac{1}{2}+\frac{3}{2}(2\alpha-1)^2].\notag
\label{54}
\end{eqnarray}
The asymptotic form of the photon wave function $\phi(u)$ at the renormalization scale ($\mu=1\textrm{GeV}^{2}$) is defined as 
\begin{eqnarray}
\phi(u)=\chi(\mu)u(1-u),
\label{55}
\end{eqnarray}
where $\chi(\mu)$ is the magnetic susceptibility. Also $F_{\mu\nu}$ is the electromagnetic field strength tensor that is defined as

\begin{eqnarray}
F_{\mu\nu}(x) &=& -i(\varepsilon_{\mu}q_{\nu}-\varepsilon_{\nu}q_{\mu})e^{iqx},\notag \\
\widetilde{F}_{\mu\nu}(x) &=& \frac{1}{2}\varepsilon_{\mu\nu\alpha\beta}F_{\alpha\beta}(x).
\label{57}
\end{eqnarray}

Using the $\gamma$-matrix identities and inserting the above PDA's in the correlation function (\ref{46}), we find the following formulas for the relevant invariant structures:

\begin{eqnarray}
\Pi_{1(f)}^{(D_s)}(p,q) &=& \frac{Q_s}{4 \left(m_c^2-p^2\right)^3}\int_{0}^{1} du \Big[\textbf{A}(u)\langle\bar{s}s\rangle \left(3 m_c^2+p^2\right)
\notag \\ && -2 \left(m_c^2-p^2\right) \left(f_{3\gamma} \overline{\psi}^{(A)}(u) m_c+ 2\langle\bar{s}s\rangle \phi(u)(5 m_c^2- p^2 ) \right)\Big],
\label{58}
\end{eqnarray}
and
\begin{eqnarray}
\Pi_{2(f)}^{(D_s)}(p,q) &=& \frac{Q_s}{4 \left(m_c^2-p^2\right)^3}\int_{0}^{1} du \Big[\textbf{A}(u)\langle\bar{s}s\rangle \left(3 m_c^2+p^2\right)+2\textbf{B}(u) \langle\bar{s}s\rangle \left(3 m_c^2- p^2 \langle\bar{s}s\rangle\right)
\notag \\ && -4 \left(m_c^2-p^2\right) \left(f_{3\gamma} \overline{\psi}^{(V)}(u) m_c + \langle\bar{s}s\rangle \phi(u) (5m_c^2-p^2) \right) \Big].
\label{59}
\end{eqnarray}

Finally, by matching the QCD and hadronic parts of the correlation function, and performing the Borel transformation that transforms $p^2\rightarrow M_B^2$, we can calculate the the final result for the $D_s$ meson transition form factors arising from the contribution of both perturbative and non-perturbative parts as follows:

\begin{eqnarray}
F_{(V,A)}^{(D_s)}(Q^2)=\frac{\left(m_c+m_s\right)}{e f_{D_s} m_{D_s}} e^{m_{D_s}^2/M_B^2} \widehat{B}_{p^2}\Big\{\int_{(m_{s}+m_{c})^2}^{t_{0}}dt\frac{\rho_{1,2}(t,Q^2)}{t-p^2}+\Pi_{(1,2)(c+d+e+f)}^{(D_s)}\Big\},
\label{60}
\end{eqnarray}
where $V$ and $A$ in the left hand side correspond to 1 and 2 in the right hand side, respectively.

The final part of this section is devoted to the calculation of the transition form factors for $D^*_s$ and $D_{s1}$ mesons. As these mesons are vector and axial-vector, the Lorentz decomposition of their correlation functions and their invariant structures are the same as that of vector and axial vector $B_c$ mesons (c.f. Eqs.(\ref{15}) and (\ref{16})). 

As $D_s$ meson, the invariant structures $\Pi_{1}$ and $\Pi_{2}$ for $D^*_s$ and $D_{s1}$ mesons consist of both perturbative and non-perturbative parts.

The steps to calculate the perturbative parts for both $D^*_s$ and $D_{s1}$ mesons are the same as that of $B_c$ in the previous section. So we only write the final expressions for the spectral densities for $D^*_s$ meson that are

\begin{eqnarray}
\rho_1^{(D^*_s)} (t,Q^2) &&= \frac{-e N_c t^2}{8 \pi ^2 \left(t-Q^2\right)^3}  \Bigg\{ Q_s \Bigg[ \lambda  \Bigg(Q^2 \Big[\alpha (\alpha -7)  +\beta (5 -2\alpha + \beta)\Big]
+2 t \Big[ \beta (3 \alpha + 4 \beta)-\alpha (1-\alpha )\Big]\Bigg)
\notag \\ && + 2 \alpha   \Bigg(Q^2 (\alpha -4 \beta +2)+t \Big[\alpha(1-\alpha) + \beta (3\alpha + 6\beta -4) \Big]\Bigg) \text{Ln}\left(\frac{1+\alpha -\beta +\lambda}{1+\alpha -\beta -\lambda}\right) \Bigg]
\notag \\ && + Q_c \Bigg[  \lambda  \Bigg(Q^2 \Big[\beta (\beta -7)  +\alpha (5 -2\beta + \alpha)\Big]
+2 t \Big[ \alpha (3 \beta + 4 \alpha)-\beta (1-\beta )\Big]\Bigg)
\notag \\ && + 2 \beta   \Bigg(Q^2 (\beta -4 \alpha +2)+t \Big[\beta(1-\beta) + \alpha (3\beta + 6\alpha -4) \Big]\Bigg) \text{Ln}\left(\frac{1+\beta -\alpha +\lambda}{1+\beta -\alpha -\lambda}\right)  \Bigg]   \Bigg\},
\label{62}
\end{eqnarray}
and
\begin{eqnarray}
\rho_2^{(D^*_s)}  (t,Q^2) &&=  \frac{e N_c t^2}{8 \pi ^2 \left(t-Q^2\right)^3} \Bigg\{  Q_s \Bigg[ \lambda  \Bigg(Q^2 \Big[\alpha  (\alpha +9)-\beta  (2 \alpha -\beta +3)\Big]+2 \alpha  t (3 \alpha -7 \beta +1)\Bigg)
\notag \\ && -2 \alpha   \Bigg(Q^2 (3 \alpha -4 \beta +2)+t \Big[\alpha  (\alpha +3)-\beta  (3 \alpha +6 \beta
+4)\Big]\Bigg) \text{Ln}\left(\frac{1+\alpha -\beta +\lambda}{1+\alpha -\beta -\lambda}\right) \Bigg]
\notag \\ && + Q_c \Bigg[  \lambda  \Bigg(Q^2 \Big[\beta  (\beta +9)-\alpha  (2 \beta -\alpha +3)\Big]+2 \beta  t (3 \beta -7 \alpha +1)\Bigg)
\\ && -2 \beta   \Bigg(Q^2 (3 \beta -4 \alpha +2)+t \Big[\beta  (\beta +3)-\alpha  (3 \beta +6 \alpha
+4)\Big]\Bigg) \text{Ln}\left(\frac{1+\beta -\alpha +\lambda}{1+\beta -\alpha -\lambda}\right) \Bigg] \Bigg\}, \notag
\label{63}
\end{eqnarray}
and for $D_{s1}$ meson

\begin{eqnarray}
\rho_1^{(D_{s1})} (t,Q^2) &&=  \frac{e N_c t}{8 \pi ^2 \left(t-Q^2\right)^2}
\Bigg\{ Q_s \Bigg[\lambda  \Bigg(Q^2 (1-\alpha +\beta)-t \Big[\alpha  (\alpha -2 \beta +7)+ \beta(\beta -3) \Big] \Bigg) \Bigg] 
\notag \\ && + Q_c \Bigg[\lambda  \Bigg(Q^2 (1-\beta +\alpha)-t \Big[\beta  (\beta -2 \alpha +7)+ \alpha(\alpha -3) \Big] \Bigg) \Bigg] \Bigg\},
\label{64}
\end{eqnarray}
and
\begin{eqnarray}
\rho_2^{(D_{s1})} (t,Q^2) &&=  \frac{e N_c t}{8 \pi ^2 \left(t-Q^2\right)^2}
\Bigg\{ Q_s \Bigg[ \lambda  \Bigg(2 Q^2 (\alpha -\beta )+t \Big[1+\alpha  (\alpha +4)-\beta  (2 \alpha +\beta ) \Big] \Bigg)
\notag \\ && -2 \alpha  \Big[Q^2+t (1+2 \alpha -2 \beta)\Big]  \text{Ln}\left(\frac{1+\alpha -\beta +\lambda}{1+\alpha -\beta -\lambda}\right) \Bigg]
\notag \\ && + Q_c \Bigg[ \lambda  \Bigg(2 Q^2 (\beta -\alpha )+t \Big[1+\beta  (\beta +4)-\alpha  (2 \beta +\alpha ) \Big] \Bigg)
\notag \\ && -2 \beta  \Big[Q^2+t (1+2 \beta -2 \alpha)\Big]  \text{Ln}\left(\frac{1+\beta -\alpha +\lambda}{1+\beta -\alpha -\lambda}\right) \Bigg]
\Bigg\}.
\label{65}
\end{eqnarray}

As the non-perturbative diagrams for both $D_s^*$ and $D_{s1}$ mesons are the same as that of $D_s$, following the similar steps, the final expressions for the non-perturbative contributions of the invariant structures for $D_s^*$ meson are as follows

\begin{eqnarray}
\Pi_{1(c+d+e+f)}^{(D_s^*)}(p,q) &&=  \frac{e Q_c \langle\bar{s}s\rangle}{12 r^6 R^6} \Big[ -m_0^2 m_c \Big(24 r^2 m_c^4-6 m_c^2 \left(r^4-r^2 R^2+R^4\right)+8 r^4 R^2+3 r^2 R^4\Big)
\notag \\ && +6 \Big(-2 r^4 R^4 m_c+r^2 R^2 m_c^2 m_s \left(r^2+R^2\right)+r^2 R^2 m_c m_s^2 \left(3 r^2+R^2\right)+8 r^2 m_c^5 m_s^2
\notag \\ && -2 m_c^3 m_s^2 \left(r^4-r^2 R^2+R^4\right)+r^4 R^2 m_s
\left(r^2+R^2\right)\Big)  \Big] 
\notag \\ && -\frac{2 f_{3\gamma} Q_s }{m_c^2-p^2} \int_{0}^{1} du \psi^{(V)}(u),
\label{66}
\end{eqnarray}
and
\begin{eqnarray}
\Pi_{2(c+d+e+f)}^{(D_s^*)}(p,q) &&= \frac{e Q_c \langle\bar{s}s\rangle}{4 r^6 R^6} \Big[  m_0^2 \Big(r^2 R^2 m_c \left(4 r^2+R^2\right)+8 r^2 m_c^5-2 m_c^3 \left(r^4-r^2 R^2+R^4\right)\Big)
\notag \\ && -2 \Big(-2 r^4 R^4 m_c+r^2 R^2 m_c^2 m_s \left(r^2+R^2\right)+r^2 R^2 m_c
m_s^2 \left(3 r^2+R^2\right)+8 r^2 m_c^5 m_s^2
\notag \\ && -2 m_c^3 m_s^2 \left(r^4-r^2 R^2+R^4\right)+r^4 R^2 m_s \left(r^2+R^2\right)\Big)\Big]
\notag \\ && + \frac{m_c Q_s}{2 \left(m_c^2-p^2\right)} \int_{0}^{1} du \Big[ \textbf{A}(u) \langle\bar{s}s\rangle m_c^2
+\left(m_c^2-p^2\right) \Big(f_{3\gamma} \psi^{(A)}(u) m_c
\notag \\ && + \langle\bar{s}s\rangle [ \textbf{B}(u)-  \phi(u) (10 m_c^2+2 p^2) ] \Big)\Big],
\label{67}
\end{eqnarray}
and for the $D_{s1}$ meson we have

\begin{eqnarray}
\Pi_{1(c+d+e+f)}^{(D_{s1})}(p,q) &&= \frac{e Q_c \langle\bar{s}s\rangle}{12 r^6 R^6} \Big[ -m_0^2 m_c \Big(24 r^2 m_c^4-6 m_c^2 \left(r^4-r^2 R^2+R^4\right)+8 r^4 R^2+3 r^2 R^4\Big)
\notag \\ && + 6 \Big( r^2 R^2 m_c^2 m_s \left(r^2+R^2\right)+r^2 R^2 m_c \left(m_s^2 \left(3 r^2+R^2\right)-2 r^2 R^2\right)+8 r^2 m_c^5 m_s^2
\notag \\ && -2 m_c^3 m_s^2 \left(r^4-r^2 R^2+R^4\right)+2 r^4 R^4 m_s\Big) \Big]  +\frac{ f_{3\gamma} Q_s }{m_c^2-p^2} \int_{0}^{1} du \psi^{(V)}(u),
\label{68}
\end{eqnarray}
and
\begin{eqnarray}
\Pi_{2(c+d+e+f)}^{(D_{s1})}(p,q)  &&= \dfrac{e Q_c \langle\bar{s}s\rangle}{4 r^6 R^6} \Big[ m_0^2 \Big(r^2 R^2 m_c \left(4 r^2+R^2\right)+8 r^2 m_c^5-2 m_c^3 \left(r^4-r^2 R^2+R^4\right)\Big) 
\notag \\ && + 2 \Big(2 r^4 R^4 m_c+r^2 R^2 m_c^2 m_s \left(r^2+R^2\right)-r^2 R^2 m_c m_s^2 \left(3 r^2+R^2\right)-8 r^2 m_c^5 m_s^2 
\notag \\ && +2 m_c^3 m_s^2 \left(r^4-r^2 R^2+R^4\right)+r^4 R^2 m_s
\left(r^2+R^2\right)\Big)\Big]
\notag \\ && -\frac{f_{3\gamma} Q_s m_c^2  }{\left(m_c^2 -p^2\right)^2} \int_{0}^{1} du \psi^{(A)}(u).
\label{69}
\end{eqnarray}

Putting all contributions together and performing the Borel transformation, the final expressions for transition form factors for $D^*_s$ (and also the same for $D_{s1}$) would be:

\begin{eqnarray}
F_{(V,A)}^{(D^*_s)}(Q^2)= \frac{f_{D^*_s}}{e~ m_{D^*_s}} e^{m_{D^*_s}/M^2} \widehat{B}_{p^2} \Big\{ \int_{(m_s + m_c)^2}^{t_0} dt \dfrac{\rho^{(D^*_s)}_{(1,2)}(t,Q^2)}{t-p^2} +\Pi_{(1,2)(c+d+e+f)}^{(D^*_s)}\Big\}.
\label{70}
\end{eqnarray}

\section{NUMERICAL ANALYSIS\label{Bound}}
In this section we calculate the decay rate and branching ratios of transitions under consideration by using the fit functions of the form factors. To this end, we use the set of input parameters shown in table (\ref{Table1}). 
For the threshold parameters we use $s_{0}^{(B_c^{(V)})}= 45 GeV^{2}$, $s_{0}^{(B_c^{(A)})}= 54 GeV^{2}$, $t_{0}^{(D_s)}=6.5 GeV^{2}$, $t_{0}^{(D^*_s)}=8 GeV^{2}$ and  $t_{0}^{(D_{s1})}=8 GeV^{2}$\cite{Wang:2012kw,Aliev1, Khosravi,Ref27}. 
To evaluate the transition form factors, we also need to find the regions for the auxilary Borel parametes, in order that the form factors to be practically independent of them. The suitable ranges for the Borel parameters are $10 GeV^2<M^2_{B_c}<15 GeV^2$, $2 GeV^2<M^2_{D_s}<5 GeV^2$, $2 GeV^2<M^2_{D^*_s}<5 GeV^2$ and $2 GeV^2<M^2_{D_{s1}}<5 GeV^2$. 

\begin{table}[ht]
	\centering
	\begin{tabular}{cccc}
		\hline \hline
		Input Parameters  &  Values   &  Input Parameters  &  Values 
		\\
		\hline \hline
		$ m_{c} $   &   $1.28 \pm 0.03$ $GeV$  & 	$ m_{s} $ &   $ 96^{+8}_{-4} $ $MeV$ \\
		$ m_{b} $   &   $4.18^{+0.04}_{-0.03} $ $GeV$  &$m_{D_{s1}}$ &  $2459.5 \pm 0.6 $ $MeV$  \\
		$m_{D^*_s}$ &  $ 2112.1 \pm 0.4 $ $MeV$ & $ m_{D_{s}}$ &  $1968.28 \pm 0.10 $ $MeV$ \\
		$m_{B_{c}^{(V)}}$ & $ 6.331 \pm 0.047 $ $GeV$\cite{Wang:2012kw} &$ m_{B_{c}^{(A)}}$ &  $ 6.737 \pm 0.056 $ $GeV$ \cite{Wang:2012kw}  \\
		
		$ G_{F} $  &  $ 1.166\times 10^{-5} $ $GeV^{-2}$ &	$ \alpha_{em} $ & $ 1/137 $ \\
		
		$ |V_{cs}|$ & $ 0.995 $ & $ |V_{cb}|$  & $0.0422 $ \\
		
		$ f_{D_s} $   &    $249$ $MeV$ & $ f_{D_{s1}} $   &    $225$ $MeV$ \cite{Khosravi} \\
		
		$ f_{D^*_s} $  &   $ 266 $ $MeV$  \cite{Ref27} &$ f_{B_{c}^{(V)}} $   &    $415$ $MeV$ \cite{Wang:2012kw} \\
		$ f_{B_{c}^{(A)}} $   &    $374$ $MeV$ \cite{Wang:2012kw} & $\langle\overline{\psi}\psi|_{\mu=1GeV}\rangle$ & $-(240 \pm 10 MeV)^{3}$ \cite{Ref22}     \\
		$\langle\overline{s}s\rangle$  & $ (0.8 \pm 0.2)\langle\overline{\psi}\psi\rangle $ \cite{Ref22} & $\chi(\mu=1 GeV)$ & $0.3 GeV^{-2}$ \\
		$k$ &  $0.2$ & $k^{+}$ &  0 \\
		$\zeta_{1}$ & 0.4  & $\zeta_{1}^{+}$  &  0 \\
		$\zeta_{2}$  & 0.3  &  $\zeta_{2}^{+}$  & 0  \\
		$f_{3\gamma}$ &  $-(4 \pm 2)\times 10^{-3} GeV^{2}$  &  $\omega_{\gamma}^{A}$  & $-2.1 \pm 1.0$   \\
		$\omega_{\gamma}^{V}$ & $3.8 \pm 1.8$  &  $\tau_{B_{c}}$ &  $0.52 \times 10^{-12} s$ \cite{Ref31}  \\
		
		\hline \hline
	\end{tabular}
	\caption{The values of some input parameters used in the numerical analysis. They are mainly taken from PDG \cite{Patrignani:2016xqp}, except ones that the references are cited next to the numbers. Also the PDA's parameters are taken from \cite{DA1,DA2,Ref28}}
	\label{Table1}
\end{table}

Now, we need to find the fit function of the form factors. The general formula that is used for $D_s$, $D^*_s$ and $D_{s1}$ mesons has the generic form of

\begin{eqnarray}
f^{D_s}_{A,V}(Q^2)=\dfrac{f^{D_s}_{A,V}(0)}{1+a^{D_s}_{A,V}\frac{Q^2}{m^2_{D_s}}+b^{D_s}_{A,V}(\frac{Q^2}{m^2_{D_s}})^2},
\label{74}
\end{eqnarray} 
that the relevant form factor should be calculated at $Q^2=m^2_{B_c}$.
Also the generic form of the fit function for $B_c$ mesons is as follows:

\begin{eqnarray}
g^{B_c}_{A,V}(p^2)=\dfrac{g^{B_c}_{A,V}(0)}{1+a^{B_c}_{A,V}\frac{p^2}{m^2_{B_c}}},
\label{75}
\end{eqnarray} 
where $g^{B_c}_{A,V}(p^2)$ have to be evaluated at $p^2=m^2_{D_s}$. 
The relevant fit parameters are shown in tables \ref{Table2} and \ref{Table3}.

\begin{table}[ht]
	\centering
	\begin{tabular}{cccc}
		\hline \hline
		fit Parameters  &  Values   &  fit Parameters  &  Values 
		\\
		\hline \hline
		$F^{B_c^{(V)}}_{V}(0)$ & $0.459 \pm 0.064$  & $a^{B_c^{(V)}}_{V}$  & -1.809 \\
		
		$F^{B_c^{(V)}}_{A}(0)$ & $ 0.541 \pm 0.081 $ & $a^{B_c^{(V)}}_{A}$  & -1.808 \\
		
		$F^{B_c^{(A)}}_{V}(0)$ & $ 0.541 \pm 0.092 $ & $a^{B_c^{(A)}}_{V}$  & -2.048 \\
		
		$F^{B_c^{(A)}}_{A}(0)$ & $ 0.459 \pm 0.083 $ & $a^{B_c^{(A)}}_{A}$  & -2.049 \\
		
		\hline \hline
	\end{tabular}
	\caption{The fit parameters for vector and axial-vector $B_c$ transition form factors fit functions}
	\label{Table2}
\end{table}

\begin{table}[ht]
	\centering
	\begin{tabular}{cccccc}
		\hline \hline
		fit Parameters  &  Values   &  fit Parameters  &  Values  &  fit Parameters  &  Values
		\\
		\hline \hline
		$F^{D_s}_{V}(0)$ & $ -0.097 \pm 0.013 $ & $a^{D_s}_{V}$  & 0.154 & $b^{D_s}_{V}$  & -0.025 \\
		
		$F^{D^*_s}_{V}(0)$ & $ 1.768 \pm 0.212 $ & $a^{D^*_s}_{V}$  & 19.798 & $b^{D^*_s}_{V}$  & -2.06284 \\
		
		$F^{D_{s1}}_{V}(0)$ & $ 1.211 \pm 0.193 $ & $a^{D_{s1}}_{V}$  & 0.511 & $b^{D_{s1}}_{V}$  & 0.055 \\	
		\hline \hline
	\end{tabular}
	\caption{The fit parameters for $D_s$, $D^*_s$ and $D_{s1}$ meson transition form factors fit functions}
	\label{Table3}
\end{table}

From Eqs. (\ref{MBctoDs}) and (\ref{MBctoDss}) one can obtain the decay rates for the relevant tarnsition as

\begin{eqnarray}
\Gamma(B_c \rightarrow D_s ~ \gamma) &&=  \frac{e^2 G_F^2}{4}  V_{cb}^2 V_{cs}^2 \Big(\dfrac{m_{B_c}^2-m_{D_s}^2}{m_{B_c}^2 m_{D_s}^2}\Big)^2 
\Bigg[ f_{B_c}^2 m_{B_c}^4 \Bigg( \Big(F_{V}^{(D_s)}\Big)^2 + \Big(F_{A}^{(D_s)}\Big)^2  \Bigg)
\notag \\ && + f_{D_s}^2 m_{D_s}^4  \Bigg( \Big(F_{V}^{(B_c)}\Big)^2 + \Big(F_{A}^{(B_c)}\Big)^2  \Bigg)
\\ &&  + \dfrac{2 f_{B_c}^2 f_{D_s} m_{B_c}^2 m_{D_s}^2}{(m_{B_c}^2 - m_{D_s}^2)^2} \Bigg( 2 f_{D_s} m_{D_s}^2 \Big(2 m_{B_c}^2 - m_{D_s}^2\Big) -F_{A}^{(D_s)} \Big(m_{B_c}^2 - m_{D_s}^2 \Big) \Big(3 m_{B_c}^2 + m_{D_s}^2 \Big) \Bigg) \Bigg], \notag
\label{GBctoDs}
\end{eqnarray}
and
\begin{eqnarray}
\Gamma(B_c \rightarrow D^*_s ~ \gamma) &&= \frac{e^2 G_F^2}{4}  V_{cb}^2 V_{cs}^2 \frac{\left(m_{B_c}^2-m_{D^*_s}^2\right)^2 \left(m_{B_c}^2+m_{D^*_s}^2\right)}{m_{B_c}^6 m_{D^*_s}^6} \Bigg[ f_{B_c}^2 m_{B_c}^6 \Bigg(\Big(F_V^{(D^*_s)}\Big)^2+\Big(F_A^{(D^*_s)}\Big)^2\Bigg)
\notag \\ && + f_{D^*_s}^2 m_{D^*_s}^6 \Bigg(\Big(F_V^{(B_c)}\Big)^2+\Big(F_A^{(B_c)}\Big)^2\Bigg)  -2 f_{B_c} f_{D^*_s} m_{B_c}^3  m_{D^*_s}^3 \left(F_V^{(B_c)} F_V^{(D^*_s)} +F_A^{(B_c)} F_A^{(D^*_s)} \right)
\notag \\ && +6 f_{B_c}^2  f_{D^*_s}^2 \frac{ m_{B_c}^6 m_{D^*_s}^6}{\left(m_{B_c}^2-m_{D^*_s}^2\right)^2}\Bigg],
\label{GBctoDss}
\end{eqnarray} 
which can be used for $B_c$ to $D_{s1}$ transition either.

Finally, the numerical values of the corresponding branching ratios for these decays are obtained as follows:
\begin{eqnarray}\label{Br}
\textbf{B}(B_{c}^{(V)}\to D_{s} \gamma) &=& (7.382 \pm 2.067) \times 10^{-5}, \notag \\
\textbf{B}(B_{c}^{(V)}\to D^*_{s} \gamma) &=& (6.290 \pm 1.824) \times 10^{-5}, \notag \\
\textbf{B}(B_{c}^{(V)}\to D_{s1} \gamma) &=& (2.393 \pm 0.598) \times 10^{-5},\notag \\
\textbf{B}(B_{c}^{(A)}\to D_{s} \gamma) &=& (2.606 \pm 0.782) \times 10^{-6},  \\
\textbf{B}(B_{c}^{(A)}\to D^*_{s} \gamma) &=& (5.259 \pm 1.367) \times 10^{-5}, \notag \\
\textbf{B}(B_{c}^{(A)}\to D_{s1} \gamma) &=& (2.768 \pm 0.775) \times 10^{-5}. \notag 
\end{eqnarray}
We see that the branching ratios for the relevant tarnsitions are overall of order of $10^{-5}$ that means they can be observed at LHCb in near future.

\section{CONCLUSIONS\label{Bound}}

We have studied the radiative decays $B_c^{(A,V)} \rightarrow D_s \gamma$, $B_c^{(A,V)} \rightarrow D^*_s \gamma$ and $B_c^{(A,V)} \rightarrow D_{s1} \gamma$. To this end, first we calculated the relevant form factors entering the amplitudes defining these transitions. By fixing the auxilary parameters we found the fit functions of the form factors at these decay channels. We used them to estimate the partial decay widths as well as the branching ratios of the considered transitions.
The order of branching ratios show that these channels are accessible in the near future experiments.

\acknowledgements

The author would like to thank K. Azizi for useful discussions and valuable comments. Also the warm hospitality of Institute for Research in Fundamental Sciences (IPM) is appreciated.

\end{document}